 \documentstyle[prl,aps,epsfig]{revtex} 

\begin{document}

 \twocolumn[\hsize\textwidth\columnwidth\hsize\csname@twocolumnfalse%
 \endcsname 

\draft


\title{ \vskip -0.5cm
         \hfill\hfil{\rm\normalsize Printed on \today}\\
Electric Polarization of Heteropolar Nanotubes \\
as a  Geometric Phase }

\author{ E. J. Mele$^{1}$
and Petr Kr\'al$^{2}$}

\address{$^{1}$ Department of Physics, Laboratory for Research on the
Structure of Matter,\\
University of Pennsylvania, Philadelphia, Pennsylvania 19104}

\address{$^{2}$ Department of Chemical Physics, Weizmann Institute of Science,
         76100 Rehovot, Israel}

\date{Received \hspace{3.0cm}}

\maketitle


\begin{abstract}
The three-fold symmetry of planar boron nitride, the III-V analog
to graphene, prohibits an electric polarization in its ground
state, but this symmetry is broken when the sheet is wrapped to
form a BN nanotube. We show that this leads to an electric
polarization along the nanotube axis which is controlled by the
quantum mechanical boundary conditions on its electronic states
around the tube circumference. Thus the macroscopic dipole moment
has an {\it intrinsically nonlocal quantum} mechanical origin from the
wrapped dimension. We formulate this novel phenomenon using the
Berry's phase approach and discuss its experimental consequences.
\end{abstract}




 ] 


\narrowtext

Physical properties of materials at the nanoscale can differ
dramatically from their bulk counterparts. This is especially
evident in the electronic properties, since the quantum behavior
of electrons on this scale is sensitive to the size, shape and
symmetry of the sample. Recent discovery of carbon nanotubes
\cite{Iijima91} provides a striking example, where metallic or
semiconducting tubes of identical compositions have only slightly
different radii \cite{ham,mint,sait}. Layered BN provides a III-V
analog to these materials; it can be formed in single and
multiwall nanotubes that have the same Bravais lattice as their
graphene counterparts, but with {\it inequivalent} atomic species
on its two sublattices \cite{chop,blase}.

Here we show that the broken sublattice symmetry produces a macroscopic
electric polarization in BN nanotubes, dependent on their topology.
Remarkably, this ground state polarization is an {\it intrinsically
nonlocal quantum} effect that cannot be described by a
classical theory. The sign and size of the longitudinal
polarization of the heteropolar tube are controlled by the
boundary conditions on its electronic wave functions along its
{\it wrapped} compact dimensions. We analyze this novel
phenomenon by developing a quantum theory of the nanotube polarization in
terms of a geometric phase \cite{berry,ksv,resta}.

A natural description of the electronic properties of this system
is developed from an expansion of the tight binding Hamiltonian
for the $\pi$ electrons at small wavevectors $q$ around the $K$
and $K'$ points at the corners of the Brillouin zone of the 2D
hexagonal lattice \cite{km,krm}. Introducing index $\alpha = \pm
1$ for these points then leads to the long wavelength Hamiltonians
\cite{km,krm,sem}
\begin{equation}
H_{\alpha}(q, \delta, \Delta) =  \alpha \hbar v_F q \, \sigma_x +
\delta_{\alpha} \, \sigma_y + \Delta \, \sigma_z\ ,
\label{H}
\end{equation}
where $\sigma_{\mu}$ are the $2 \times 2$ Pauli matrices, $q$ is
the wavevector along the tube axis, $v_F$ is the Fermi velocity
and $\Delta$ is a site diagonal potential that distinguishes the B
and N sites. The crucial parameter $\delta_{\alpha} = \alpha
\hbar v_F q_{\perp}$ is a contribution to the electronic gap that
comes from the quantized crystal momentum $q_{\perp}$ along its
azimuthal direction.

We can interpret Eqn.~(\ref{H}) as the Hamiltonian for an
effective spin $1/2$ particle interacting with a ``magnetic
field" defined by the parameters $(q,\, \delta,\, \Delta)$.
$H_{\alpha}$ can be diagonalized by the rotation $U_{\alpha} (q,
\delta, \Delta) = \exp( i \hat \Omega_{\alpha} \cdot \sigma /2)$,
which aligns the vector
\begin{equation}
\hat \Omega_{\alpha} =(\alpha q, \delta_{\alpha},
\Delta)/\sqrt{q^2 + \delta_{\alpha}^2 + \Delta^2}\ \label{OM}
\end{equation}
(units $\hbar v_F = 1$) in the $z$ direction. The valence band
eigenfunctions $v_{\alpha}$ of Eqn.~(\ref{H}), relevant here, have
energies $E_{\alpha} (q) = -\sqrt{q^2 + \delta_{\alpha}^2 +
\Delta^2}$. They are obtained by the rotation of the down spinor
$v_0 = (0,1)$ according to $v_{\alpha}(q,\delta,\Delta) = \exp( i
\hat \Omega_{\alpha} \cdot \sigma /2) \cdot v_0$, and correspond
to the spin down state quantized along the local  $\hat
\Omega_{\alpha}$ axis.

The electric polarization of an extended system is ill defined as
an intensive quantity, but its {\it changes} are well defined
\cite{ksv,resta}. For an electronic Hamiltonian containing a
control parameter {\it H}($\lambda$), the difference in
polarization $\Delta p$ between  initial and final states at
$\lambda_i$ and $\lambda_f$ can be obtained by integrating the
differential changes in $p$ as one adiabatically varies the
control parameter $\lambda$
\begin{equation}
\Delta p = \int_{\lambda_i}^{\lambda_f} d \lambda \, \frac
{\partial p}{\partial \lambda}\ , \label{Dp}
\end{equation}
We evaluate $\Delta p$ on the path from the C nanotube, which is
nonpolar by symmetry, to the heteropolar BN nanotube which can
have a macroscopic dipole moment. We take as a control parameter
the antisymmetric site diagonal potential $\Delta$, where
$\Delta_i = 0$ in the C nanotube. Therefore, integrating
Eqn.~(\ref{Dp}) over the range $0 < \Delta < \Delta_f$, gives an
expression for the electric polarization of the heteropolar BN
nanotube with the ionic strength $\Delta_f$.

Equation (\ref{Dp}) can be evaluated by studying the variation of
$-i\, \langle v_{\alpha} |\partial/\partial q | v_{\alpha}
\rangle$ as a function of $\Delta$. Summing over $\alpha$ and
wave vectors $q$, we find
\begin{eqnarray}
\frac{p}{e} & = & \frac{1}{2 \pi i} \sum_{\alpha} \int_0^{\Delta_f} d
\Delta \int_{- \pi}^{\pi}  dq
\nonumber \\
& \times &
\left( \langle \frac{\partial
v_{\alpha}}{\partial \Delta} | \frac{\partial
v_{\alpha}}{\partial q} \rangle - \langle \frac{\partial
v_{\alpha}}{\partial q} | \frac{\partial v_{\alpha}}{\partial
\Delta} \rangle \right)\ .
\label{POL1}
\end{eqnarray}
The valence states $v_{\alpha}$ adiabatically follow the direction of
the vector $-\hat \Omega_{\alpha}$, so that  $\partial
v_{\alpha}/\partial q = (-i/2) (\partial (\hat \Omega \cdot \sigma
)/  \partial q) v_{\alpha}$ and $\partial v_{\alpha}/\partial
\Delta = (-i/2) (\partial (\hat \Omega \cdot \sigma) /
\partial \Delta) v_{\alpha}$. Therefore, the polarization is
\begin{equation}
\frac{p}{e} = \frac{1}{4 \pi} \sum_{\alpha} \int_0^{\Delta_f} d
\Delta \int_{- \pi}^{\pi}  dq \, \frac{\langle v_{\alpha}|
\sigma_y |v_{\alpha} \rangle}{q^2 +\delta^2 +\Delta^2}\ .
\label{POL2}
\end{equation}

\begin{figure}[htbp]
\epsfxsize=3.0in
       \centerline{\epsfbox[42 164 548 633]{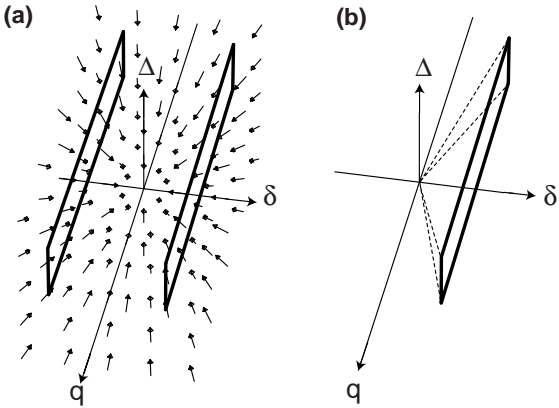}}
         \vspace{5mm}
\caption{(a) The electric polarization is the flux of a vector
field linking through two rectangular loops in the space of
Hamiltonian control parameters.  (b) The flux through each
rectangle is proportional to the solid angle subtended by the
loop, which is also swept out by the vector $\hat \Omega$ when
transported around the perimeter of the loop. }
\label{FIG1}
\end{figure}

Equation (\ref{POL2}) has a simple geometric interpretation. The
effective ``spin" representation for each valence state
$v_{\alpha}$ defines a unit vector field directed inward along
the radial direction at each point ($q,\delta,\Delta$) in the
parameter space for ${\cal H}_{\alpha}$. The two surface integrals
in (\ref{POL2}) give the flux of a radial {\it inverse square}
field in this parameter space that links through the two
rectangular loops, shown in Fig.~\ref{FIG1}a. These are located at
$\delta_{\alpha= \pm 1}=\pm \hbar v_F \, q_{\perp}$, oriented
parallel to the $q-\Delta$ plane and extending from $-\pi < q
<\pi$ and $0<\Delta<\Delta_f$.
This flux is related by Stokes' theorem to the line integral of
an effective vector potential $\langle v_\alpha | \partial_{q(\Delta)}
|v_\alpha \rangle$ around the loop perimeters, shown in Fig.~\ref{FIG1}b.
Thus, the electric polarization is proportional to the
solid angle that is ``swept out" by the effective spin as it is
adiabatically transported around the loops.

Introducing an energy cutoff at the bandwidth $W = \pi \hbar
v_F$, the double integral in Eqn.~(\ref{POL2}) gives
\begin{eqnarray}
\int_0^{\Delta_f} d \Delta \int_{- \pi}^{\pi}  &dq& \,
\frac{\alpha \delta } {(q^2 +\delta^2 +\Delta^2)^{3/2}}
\nonumber\\ &=& 2 \alpha \arctan \left( \frac{W \Delta}{\delta
\sqrt{\delta^2 + \Delta^2 +W^2}} \right). \label{POL3}
\end{eqnarray}
It is intuitively clear that by interchanging the B and N
sublattices, {\it i.e.} negating $\Delta$, the sign of $p$ should
change. Much more profound is the fact the dipole moment in
Eqn.~(\ref{POL3}) is also {\it odd} in $\delta$, which thus
reflects its essential quantum mechanical origin ``rooted in the
wrapped dimension".

The macroscopic polarization is obtained by summing the allowed
values of $\delta$ for each occupied subband in Eqn.~(\ref{POL3}).
When a BN sheet is wrapped into a tube the transverse crystal
momentum is quantized $q_{\perp,n}$, and this restricts the
allowed values of $\delta$ in the sum.  The transverse momenta
$q_{\perp,n}$ depend on the subband index, $n$  and the wrapping
vector $\vec {\cal C}_{MN} = M \vec T_1 +N \vec T_2$ of the BN
nanotube, where $\vec T_1$ and $\vec T_2$ are the primitive
translation vectors for the hexagonal lattice
\cite{ham,mint,sait,mint2}. A tube with wrapping indices $(M,N)$
has a chiral index $\nu = {\rm mod}\, (M-N,3)$ \cite{krm} and its
transverse momenta are quantized to the values $q_{\perp, n} =(2
\pi/|\vec {\cal C}_{MN}|)\, (n + \frac{1}{3}\, {\rm sgn} \,
\nu)$, so the gap parameter is $\delta_{\alpha n} = \alpha \hbar
v_F \, q_{\perp,n}$. Thus summing Eqn.~(\ref{POL3}) over subbands
finally gives the net electric polarization
\begin{equation}
\frac{p}{e} = \frac{1}{2 \pi} \sum_{ \alpha n} \alpha \arctan
\left( \frac{W \Delta}{\delta_{\alpha n} \sqrt{\delta_{\alpha
n}^2 + \Delta^2 + W^2}} \right)\ .
\label{POL4}
\end{equation}
Note that the summand is {\it odd} in $\delta_{\alpha n}$, but
{\it even} under the interchange $\alpha \leftrightarrow -\alpha$.
Thus contributions to the electric polarization from states near
the $K$ and $K'$ points are additive. However, a nonzero electric
dipole moment occurs only when the momentum distribution around
either point is odd in $\delta_{\alpha n}$. Thus the quantization
of the transverse crystal momentum is the critical feature that
controls the electric polarization of the BN nanotube.

Eqn.~(\ref{POL4}) predicts that BN nanotubes with crystal momentum
distributions even in $q_{\perp}$ are {\it unpolarized}. This
happens in the limit of the flat BN sheet, for which the
macroscopic dipole moment is obtained from an integral over all
two-dimensional wavevectors, and for tubes with chiral indices
$\nu = 0$, for which the kinematically allowed momenta occur in
$\pm \delta_{\alpha n}$ pairs. A simple example of the latter
occurs for the ``armchair" $M=N$ structures with the mirror
symmetry (or invariance under $M \leftrightarrow N$)
\cite{armzig}. Interestingly, Eqn.~(\ref{POL4}) predicts a zero
polarization even for ``nonarmchair" tubes with $\nu = 0$.

However, for tubes with $\nu \neq 0$ the $\delta \leftrightarrow
-\delta$ symmetry is broken by the {\it fractional quantization}
of the transverse crystal momentum, and consequently these tubes
have  a net electric polarization with the sign determined by
$\nu$. This is a remarkable result, as it implies that two nearby
structures $(M,N)$ and $(M+1,N)$ can have {\it opposing} electric
dipole moments, even though their structures in the tangent plane
of the tube are nearly identical.

\begin{figure}[htbp]
\epsfxsize=3.0in
 \vspace{40mm}
      \centerline{\epsfbox[191 325 436 489]{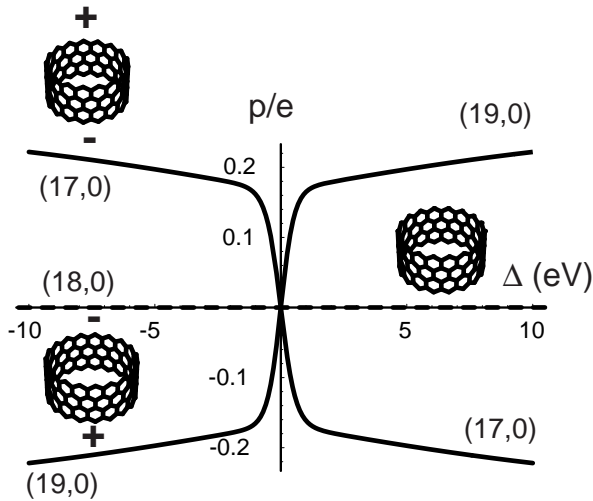}}
          \vspace{5mm}
\caption{The polarizations (dipole per unit length) for three
heteropolar zigag tubes are plotted as functions of their
ionicity parameters $\Delta$. The curves are labelled by the tube
wrapping indices $(M,N)$, and the insets give their skeleton
lattice structures. The $(18,0)$ tube has no net polarization
while its neighbors $(18 \pm 1,0)$ have nonzero electric
polarizations of opposite sign.}
\label{FIG2}
\end{figure}

A striking illustration of this effect is given in
Fig.~\ref{FIG2}, in which we plot the electric polarization (the
dipole moment per unit length in one dimension), calculated as a
function of the symmetry breaking potential $\Delta$, for three
different ``zigzag" tubes \cite{armzig} with the nearby wrapping
indices $(M,N)$ = $(17,0)$, $(18,0)$ and $(19,0)$. Skeleton
lattice structures for these three tubes are shown in the insets.
We use the following parameters representative of BN: $\hbar v_F
= 5.4\, {\rm eV \, \AA}$, $W =10 \, {\rm eV}$ and $\Delta = 2.5\,
{\rm \, eV}$, and plot the polarization in units of the
elementary charge $e$.

Figure \ref{FIG2} shows that the electric polarization is {\it
zero} for the $(18,0)$ tube, and it is {\rm inverted} for the
$(17,0)$ and $(19,0)$ tubes, despite the fact that the three
tubes have nearly identical local atomic structures in their
surface planes. This reflects the interesting fact that the dipole
moment arises as a {\it nonlocal} quantum effect that is
controlled by the continuity of the electronic wavefunctions
around the tube circumference and not any local property of the BN
lattice. In this way, the circumferential boundary conditions for
the electronic Bloch states can ultimately distinguish the three
structures by their chiral indices $\nu$.

The polarization (when nonvanishing) for BN tubes corresponds to a
dipole moment of nearly $0.5 \, D$ per unit cell. Its sign, determined
by the chiral index $\nu$,  alternates with a three sublattice period as
a function of the wrapping indices $M$ and $N$. This effect is the
heteropolar analog of three sublattice symmetry that
distinguishes conducting and semiconducting behavior of graphene
nanotubes \cite{ham,mint,sait}. Thus semiconducting graphene
nanotubes with $\nu \neq 0$ may be classified into two families
that are distinguished by the {\it sign} of their gap parameters.
This controls the sign of the polarization induced by adiabatic
introduction of the ionicity parameter $\Delta$ in
Eqn.~(\ref{Dp}).

The results derived here from the Hamiltonian (\ref{H}) give the
contribution to the electric polarization from the $\pi$ derived
electronic states. In general, one can expect additional
contributions to the electric dipole moment from lower lying
$\sigma$ bands. However, these contributions are not expected to
show a strong dependence on the chiral index and should be
smaller than the effects derived from the more polarizable $\pi$
manifold. Thus the periodic dependence of the electric
polarization on the wrapping vector arises from the phase matching
of the $\pi$-derived  Bloch states around the tube circumference.
It will be useful to quantify the relative $\pi$- and $\sigma$-
derived contributions to the electric polarization within a
complete microsopic theory.

Experimental observation of the macroscopic dipole moment of a
heteropolar tube is complicated by the possibility that extrinsic
or surface charges accumulated at the tube ends could mask the
predicted intrinsic effects. Nevertheless, the polarization $p$
in Eqn.~(\ref{POL4}) defines (modulo $e$) the surface charge that
must appear at the ends of the $\nu \neq 0$ heteropolar nanotube,
{\it independent} of the atomic structure of the termination.
Thus two-thirds of heteropolar single wall nanotubes should
contain a universal but nonintegral surface charge, determined by
the {\it bulk} electronic structure. Similarly junctions between
inequivalent tube segments are predicted to localize interface
charges $0,\pm p$ or $\pm 2p$ modulo $e$.

Multiwall heteropolar nanotubes would favor combinations which
{\it compensate} the macroscopic dipole moment.  For a double wall
tube of outer radius $a$ and inner radius $b$, one can estimate
the electrostatic contribution to the surface energy
\begin{equation}
U_s = \frac{2 z^2 e^2}{\pi |a-b|} K\Bigl(- \frac{4 ab}{(a-b)^2}\Bigr)\, ,
\end{equation}
where $ze$ is the effective bound charge accumulated at the tube
end and $K$ is the complete elliptic integral of the first kind.
Taking $z \approx 0.2$, $a \approx 10.4\,  {\rm \AA}$ and $b \approx 7.0\,
{\rm \AA}$ gives $U_s \approx 65$ meV. Note that the structure of
the double wall tubes grown under experimental conditions can be
influenced by various kinetic factors as well.

Our model can be experimentally tested by measuring the {\it
change} in polarization induced by varying control parameters in
the Hamiltonian. Of particular interest is the effect of an
elastic strain $\epsilon$ linearly coupled to the gap parameter
$\delta\rightarrow \delta+ \lambda \epsilon$. Thus the
heteropolar tube is a {\it molecular piezoelectric} with the
response constant $z^* = \partial p /\partial \epsilon = \lambda
\partial p /\partial \delta$ given by
\begin{equation}
\frac{z^*}{e} = - \frac{\lambda}{2 \pi} \frac{\Delta W (2
\delta^2 + \Delta^2 + W^2)}{(\delta^2 + \Delta^2)(\delta^2 + W^2)
\sqrt{\delta^2 + \Delta^2 + W^2}}\ . \label{PIEZO}
\end{equation}
The gap parameter $\delta$ can be linearly coupled to long
wavelength compression, extension or torsion of the tube depending
on the wrapping indices \cite{anan}. These strains are the tube
analogs to the orthorhombic and shear strains of an isolated
graphene sheet.

\begin{figure}[htbp]
\epsfxsize=3.5in
      \centerline{\epsfbox[165 309 447 514]{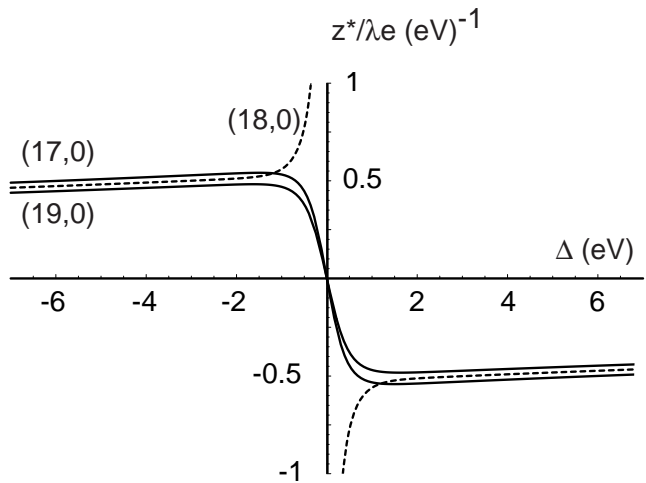}}
          \vspace{5mm}
\caption{The piezoelectric constants for the heteropolar zigzag
$(17,0)$,$(18,0)$ and $(19,0)$ tubes are plotted as a function of
the ionicity parameter $\Delta$.} \label{FIG3}
\end{figure}

In Fig.~\ref{FIG3} we plot this piezoelectric constant as a
function of the ionicity parameter $\Delta$ for the heteropolar
$(17,0)$, $(18,0)$ and $(19,0)$ nanotubes. Although the static
dipole moments $p/e$ for the $(17,0)$ and $(19,0)$ tubes have
{\it opposite} signs, their piezoelectric constants $z^*/e$ have
the {\it same} signs. This is because the couplings of the gap to
strain are nearly identical for the two structures. The
piezoelectric constants should be large for the BN tubes, since
we expect that $\lambda \approx 1 \, {\rm eV}$, similar to carbon
nanotubes. Note that the unstrained $(18,0)$ tube has no static
dipole moment but it has an exceptionally large piezoelectric
constant, that diverges proportional to $1/\Delta$ as $\Delta
\rightarrow 0$. This occurs because $\Delta=0$ marks a critical
point, where the gap vanishes for tubes with chiral index $\nu =
0$.  Figures \ref{FIG2}-\ref{FIG3} also demonstrate that BN
realizes the ``strongly ionic" limit of this problem, in which the
static dipole and its piezoelectric coefficient are essentially
saturated.

The electric dipole moments of heteropolar tubes provide crucial
information for understanding their structures, elementary
excitations and phase behavior \cite{louiseau,demzett}. They are
related, for example, with an assortment of new photogalvanic
effects \cite{krm}. It is inevitable that other physical
phenomena and applications will derive from the unique properties
of nanotubes with controllable electric polarization. This
macroscopic phenomenon also provides a beautiful and
experimentally relevant illustration of the important role of the
geometric phase in quantum mechanics.

\vspace{5mm} \noindent The work at the University of Pennsylvania
was supported by the Department of Energy under Grants
DE-FG02-01ER45118 and through the NSF through the Penn MRSEC under
grant DMR 00-79909. We thank J. Bernholc and C.L. Kane for useful
discussions.  PK would like to acknowledge M. Shapiro for support.


\begin{references}
\bibitem{armzig} An armchair tube has one-third of the BN bonds
oriented exactly along the azimuthal direction. The name is
derived from the atomic motif repeated around the tube
circumference which resembles an armchair. On a zigag tube a
chain of BN bonds ``zigzags" around the tube circumference.
\end{references}
\end{document}